\documentclass[11pt,a4paper]{article}
\usepackage{amsfonts,graphicx}

\newtheorem{thm}{Theorem}
\newtheorem{lem}{Lemma}
\newtheorem{cor}{Corollary}
\newtheorem{prop}{Proposition}
\newtheorem{defn}{Definition}
\newtheorem{eg}{Example}

\newcommand{\cL}{{\cal L}}
\newcommand{\1}{\mathbb{I}}
\newcommand{\U}{{\sf U}}

\newcommand{\Lp}[1]{\Pi_{#1}}

\newcommand{\G}{\Gamma}

\newcommand{\js}[1]{f_{+{#1}}}
\newcommand{\jsb}[1]{f_{-{#1}}}
\newcommand{\jsd}[1]{f_{\pm{#1}}}

\renewcommand{\th}[1]{\theta_{#1}}
\newcommand{\ph}[1]{\phi_{#1}}
\newcommand{\swz}[1]{\xi_{#1}}
\newcommand{\swp}[1]{\psi_{#1}}
\newcommand{\swpb}[1]{\bar{\psi}_{#1}}
\newcommand{\swq}[1]{\chi_{#1}}

\newcommand{\R}{\mathbb{R}}
\newcommand{\C}{\mathbb{C}}

\newcommand{\supp}{\mbox{supp}}

\title{Hermitian symplectic geometry and the factorisation of the scattering matrix on graphs\thanks{AMS: Primary 34B45, 34L40; Secondary 47A40, 81U20}}
\author{M. Harmer}
\date{}

\begin{document}

\maketitle 

\begin{abstract}
Hermitian symplectic spaces provide a natural framework for the
extension theory of symmetric operators. Here we show that hermitian
symplectic spaces may also be used to describe the solution to the
factorisation problem for the scattering matrix on a graph, ie. we derive
a formula for the scattering matrix of a graph in terms of the scattering
matrices of its subgraphs. The solution of this problem is shown
to be given by the intersection of a Lagrange plane and a coisotropic
subspace which, in an appropriate hermitian symplectic space, forms a
new Lagrange plane. The scattering matrix is given by a distinguished
basis to the Lagrange plane. \\
Using our construction we are also able to give a simple proof of the
unitarity of the scattering matrix as well as provide a characterisation
of the discrete eigenvalues embedded in the continuous spectrum. 
\end{abstract}

\section{Introduction}

As is well known hermitian symplectic spaces provide a natural framework
for the description of the extensions of symmetric operators
\cite{Pav,Kost:Sch,Har4}. Here we discuss another possible application
of hermitian symplectic spaces, viz. for the problem of the
factorisation of the scattering matrix for the Schr\"{o}dinger operator on
a graph. Using the fact that the Wronskian is a
hermitian symplectic form we construct a hermitian symplectic space of
solutions on the rays of the graph. We use the term asymptotic hermitian
symplectic space in analogy with the asymptotic
symplectic space introduced by Novikov \cite{Nov3} in the case
of the discrete Schr\"{o}dinger operator on a graph (in \cite{Nov3} the
Wronskian is defined so that it is a symplectic form). The value of this
construction lies in the fact that the generalised eigenspace of a
self-adjoint Schr\"{o}dinger operator forms a Lagrange plane in this
space. This allows us to easily prove the unitarity of the scattering
matrix on the real axis in the spectral plane. Furthermore we show that
the scattering matrix plays the r\^{o}le of the unitary matrix which
parameterises Lagrange planes \cite{Har4}. \\  
We also use this construction to consider the factorisation problem for
graphs. That is, we find a composition rule whereby the scattering matrix
of a graph can be written in terms of the scattering matrices of its
subgraphs. This has already been considered in the two papers  by
Kostrykin and Schrader \cite{Kost:Sch,Kost:Sch1} in the case of the
Laplacean on a graph. We present a substantially different approach,
based on the properties of the asymptotic hermitian symplectic space, to
what is essentialy the same problem, the factorisation of the
Schr\"{o}dinger operator on graphs.  Using the asymptoic hermitian
symplectic space we can express in a simple way the generalised
eigenspace  of a graph in terms of the generalised eigenspaces of the
subgraphs. As the scattering matrix is defined by a distinguished basis
for these eigenspaces this, in effect, provides us with a composition
rule for the scattering matrix. In practice, however, we need some linear
algebra (simplified by our description of the generalised eigenspace) to
write the composition rule explicitly in terms of the scattering
matrices. \\ 
Both our approach, and the approach used by Kostrykin and
Schrader, give the same answer. However we believe that our approach is
sufficiently novel to provide some new insights. Our description of the
composition rule also reveals a characterisation of the discrete
eigenvalues embedded in the continuous spectrum of the non-compact graph.

\section{The hermitian symplectic space of asymptotic solutions}

Here we study a connected non-compact graph, $\G$. We assume
that $\G$ consists of a compact part, $\G_c$, composed of $p$
finite  interior edges. Attached to arbitrary vertices of the compact
part are $n$  semi-infinite rays. Both $p$ and $n$ are finite.
Functions on the graph are represented by elements of the Hilbert space
\begin{displaymath}
H(\G)=\oplus^{n}_{i=1}L^{2}([0,\infty))\oplus^{p}_{j=1}L^{2}([0,a_{j}])
\end{displaymath}
where the $a_{j}$ are the lengths of the interior edges.
The elements of $H(\G)$ are $n+p$-dimensional vector functions and
the inner product on $H(\G)$ is
\begin{displaymath}
(\phi,\psi)_{\G}=\sum^{n}_{i=1}(\phi_i,\psi_i)_{L^{2}([0,\infty))}+
\sum^{p}_{j=1}(\phi_{n+j},\psi_{n+j})_{L^{2}([0,a_{j}])}
\end{displaymath}
where $\phi_i$ are the components of $\phi\in H(\G)$. \\
Let us consider the symmetric Schr\"{o}dinger operator,
$\cL_{0}$ in $H(\G)$ 
\begin{displaymath}
\cL_{0}\psi_i\equiv -\frac{d^2\psi_i}{dx_i^2}+q_i \psi_i \qquad
\mbox{for}\;i=1,\ldots ,n+p ,
\end{displaymath}
with domain consisting of the smooth functions 
with compact suppport in the open intervals
\begin{displaymath}
D(\cL_{0})=\oplus^{n}_{i=1}C^{\infty}_{0}([0,\infty))\oplus^{p}_{j=1}
C^{\infty}_{0}([0,a_{j}]) .
\end{displaymath}
The potentials $q_i$ are supposed to be continuous real valued functions
which are integrable with finite first moment,
\begin{equation}\label{bcndxx}
\int_{\G_i} (1 + x) \vert q_i(x) \vert dx < \infty ,
\end{equation}
where $\G_i=[0,\infty ]$ or $[0, a_i ]$.
It is easy to see that the deficiency indices of $\cL_{0}$ are 
$(n+2p,n+2p)$. Consequently we may parameterise the self-adjoint
extensions of $\cL_{0}$ by unitary matrices $\U(n+2p)$ or, for separated
boundary conditions, by $\U(d(1)) \oplus \cdots \oplus \U(d(m))$ where
$d(i)$ is the degree of the $i$-th vertex of $\G$. \\
We construct a hermitian symplectic space the elements of which are
solutions on the rays of the graph. This construction follows an analogous
construction by Novikov \cite{Nov3} for the discrete Schr\"{o}dinger
operator on graphs. Let $\psi^{\prime}$ denote the derivative of $\psi$
with respect to $x$.
\begin{defn}
The two-form $\langle\cdot,\cdot\rangle$, defined on functions on the 
{\em rays} of the graph
\begin{equation}\label{sym2}
\langle\phi,\psi\rangle \equiv\sum^{n}_{i=1}
\left[ \bar{\phi}_{i} \psi^{\prime}_{i}-\bar{\phi}^{\prime}_{i}
\psi_{i} \right] (x_i), \; x_i\in [0,\infty) 
\end{equation}
is a hermitian symplectic form.
\end{defn}
As it stands this form is not well defined, it depends on the
points $x_i$ chosen on each of the rays. If, however, we consider the set
of generalised eigenfunctions of $\cL^{\star}_{0}$ on the rays for
spectral parameter $\lambda$, 
\[
H_{2n}(\G,\cL_{0},\lambda)=\left\{ \phi\in\oplus^{n}_{i=1}C^2_{\mbox{loc}}
([0,\infty)); \; -\frac{d^2\phi_i}{dx^2_i}+q_i \phi_i=\lambda\psi_i
\right\} 
\]
we see that the form is independent of the $x_i$---due to the
constance of the Wronskian. Obviously $H_{2n}(\G,\cL_{0},\lambda)$ is a 
$2n$-dimensional vector space. We note in particluar that the functions
from $H_{2n}(\G,\cL_{0},\lambda)$ do not obey any specific boundary
conditions at the vertices. Below we may assume that the graph,
$\G$, and the potentials, $\cL_0$, are given and simply write
$H_{2n}(\lambda)$ or $H_{2n}$.
\begin{prop}
The vector space of generalised eigenfunctions on the rays 
$H_{2n}(\G,\cL_{0},\lambda)$ for real $\lambda$ equipped with the hermitian
symplectic form (\ref{sym2}) is a hermitian symplectic
space, called the asymptotic hermitian symplectic space. 
\end{prop}
To prove this statement we only need show that the form is nondegenerate
\cite{Har4} which is easy to see if we consider the basis of standard
solutions $\{\th{i} ,\ph{i} \}^n$, ie. the solutions which satisfy the
boundary conditions
\begin{eqnarray*}
\left. \th{i} \right|_0 = 1, \qquad \left. \th{i,x} \right|_0
= 0  \\
 \left. \ph{i} \right|_0 = 0, \qquad \left. \ph{i,x} \right|_0
= 1 
\end{eqnarray*}
on ray $i$ and are zero on the other rays. Furthermore we note that
this basis is canonical, 
\begin{eqnarray*}
\langle\th{i},\ph{j}\rangle = \delta_{ij} 
= -\langle \ph{i},\th{j} \rangle \\
\langle \th{i},\th{j}\rangle = 0 
= \langle\ph{i},\ph{j}\rangle 
\end{eqnarray*}
so $H_{2n}$ is a canonical hermitian symplectic space \cite{Har4} (for
brevity we just use the term hermitian symplectic space here). \\
More interesting are bases constructed from the Jost solutions. We
denote by $\jsd{,j}\in H_{2n}(\lambda)$ the elements which are 
zero on all the rays except the $j$-th where they coincide with the
Jost solution with asymptotic behaviour 
\[
  \jsd{,j} \simeq e^{\pm ikx_j} ,
\]
for $x$ large and where $\lambda=k^2$.
The fact that our two-form is
defined using complex-conjugation (The operation
$f^{\dagger}(k)=\bar{f}(\bar{k})$ appears to be more natural in the
definition of the Wronskian \cite{Har,Har4}. As we will mainly be
considering $\lambda >0$ or $k\in\R$ here we will not use this.)
complicates the evaluation of it on the Jost solutions. We have
\begin{equation}\label{Jst}
\langle \js{,i},\js{,j}\rangle= 2ik\delta_{ij} 
=-\langle \jsb{,i},\jsb{,j}\rangle \\
\langle \js{,i},\jsb{,j}\rangle = 0 
=\langle \jsb{,i},\js{,j}\rangle 
\end{equation}
but only for $\lambda >0$ or $k\in\R$. \\ 
We also construct a canonical basis using the
Jost solutions; consider
\[
\swp{0,j} = \frac{\js{,j}+\jsb{,j}}{2}, \qquad
\swq{0,j} = \frac{\js{,j}-\jsb{,j}}{2ik} 
\]
where $j=1,\ldots ,n$. It is easy to see that this is a canonical
basis, for all real $\lambda$ (unlike the relations for the Jost
solutions). \\

Now let us suppose that we have defined $\cL$, a self-adjoint extension of
$\cL_0$, on the graph. The generalised eigenfunctions of $\cL$
are (not necessarily square integrable) solutions of the
eigenvalue equation $\cL\phi =\lambda\phi$. We emphasise that such a
$\phi$ is defined on the whole of $\G$ (not just the rays) and obeys some
self-adjoint boundary conditions at the nodes of $\G$. By considering 
the restriction to the rays of the graph, a generalised eigenfunction of
$\cL$ may be thought of as an element of $H_{2n}(\lambda)$.
\begin{lem}\label{l2}
Given a self adjoint extension $\cL$, the generalised eigenspace of
$\cL$ at some real $\lambda$ forms an isotropic subspace in 
$H_{2n}(\lambda)$. 
\end{lem}
{\it Proof:} We formally consider the boundary
form of generalised eigenfunctions $\phi$ and $\psi$
\begin{equation}\label{bf2}
(\cL\phi,\psi)_{\G}-(\phi,\cL\psi)_{\G} = \sum^{n}_{i=1}
\left. \left[ \bar{\phi}_i \psi^{\prime}_{i}
- \bar{\phi}^{\prime}_i \psi_i \right]\rule{0mm}{4.5mm}\right|_{0}-
\sum^{p}_{j=1}\left. \left[ \bar{\phi}_{n+j} \psi^{\prime}_{n+j}
-\bar{\phi}^{\prime}_{n+j} \psi_{n+j} \right]
\rule{0mm}{4.5mm}\right|^{a_j}_{0} .
\end{equation}
The self-adjoint boundary conditions are described by the vanishing of this
form. Furthermore, the second sum on the right hand side vanishes by the
constancy of the Wronskian on the edges so we are left with
\begin{equation}\label{sym1}
\sum^{n}_{i=1}
\left. \left[\bar{\phi}_{i} \psi^{\prime}_{i}-\bar{\phi}^{\prime}_{i}
\psi_{i}\right] \rule{0mm}{4.5mm} \right|_{0} =
\langle\phi,\psi\rangle = 0 .
\end{equation}
This completes the proof. \hspace*{\fill} $\Box$ \\

The analogous statement for the discrete operator is proved in theorem 3
of \cite{Nov3}. In fact Novikov shows in this theorem that the eigenspaces
form Lagrange planes for any complex value of
$\lambda$. In our case, the fact that the generalised eigenspaces form
Lagrange planes for any real $\lambda$ is a simple corollary of the
following lemma:
\begin{lem}\label{l1}
Given a self-adjoint extension $\cL$, the vector space of generalised
eigenfunctions of $\cL$ at real eigenvalue $\lambda$ and with support on
the rays of the graph is $n$-dimensional.
\end{lem}
{\it Proof:} Let us consider the boundary form on $\G$, 
equation (\ref{bf2}). We know that this defines
a nondegenerate hermitian symplectic form in the $2(n+2p)$-dimensional
space of boundary values and hence a hermitian symplectic space which we
denote, in the proof of this lemma, as $H_{2(n+2p)}$
\cite{Kost:Sch,Har4}---we emphasise that we are considering the space of
boundary values, not the asymptotic hermitian symplectic space defined
above. It is clear that the self-adjoint boundary conditions are
associated with 
$(n+2p)$-dimensional Lagrange planes in this space. Let us denote by $P$ 
the $(n+2p)$-dimensional Lagrange plane in $H_{2(n+2p)}$ associated with
our chosen self-adjoint $\cL$. \\ 
Now let us consider an arbitrary interior edge indexed by $i$ of length
$a$. This edge is identified with the interval $[0,a]$. We say that a
boundary condition $\psi\in H_{2(n+2p)}$ matches on this edge if 
\[
\left( \begin{array}{c}
 \left.\psi_i\right|_a  \\
  \left.\psi^{\prime}_i\right|_a    
\end{array} \right)=
\left( \begin{array}{cc}
 \left.\th{i}\right|_a &  \left.\ph{i}\right|_a  \\
  \left.\th{i}^{\prime}\right|_a &  \left.\ph{i}^{\prime}\right|_a 
\end{array} \right)
\left( \begin{array}{c}
  \left.\psi_i\right|_0  \\
  \left.\psi^{\prime}_i\right|_0
\end{array} \right) .
\]
Here $\left.\psi_i\right|_0$ and $\left.\psi^{\prime}_i\right|_0$ are the
components of $\psi\in H_{2(n+2p)}$ corresponding to one endpoint of edge
$i$, $\left.\psi_i\right|_a$ and $\left.\psi^{\prime}_i\right|_a$ are the
components of $\psi\in H_{2(n+2p)}$ corresponding to the other endpoint of
edge $i$, $\ph{i}(\lambda)$ and $\th{i}(\lambda)$ are the standard
solutions on edge $i$ and $\lambda$ is fixed in the hypothesis. \\
It is clear that the boundary conditions on edge $i$ match iff there is a
solution of $(\cL-\lambda)f=0$ on $i$, viz.
\[
f (x,\lambda) = \left.\psi_i\right|_0 \th{i} (x,\lambda) + 
\left.\psi^{\prime}_i\right|_0 \ph{i} (x,\lambda) ,
\]
whose boundary values at the ends of edge $i$ are the same as the relevant
components of $\psi\in H_{2(n+2p)}$. \\ 
The set of boundary conditions matching on all $p$ interior edges of $\G$
and with support only on these interior edges form an isotropic subspace
in $H_{2(n+2p)}$ which we denote by $N$. This fact is equivalent to the
fact that the Wronskian of two generalised eigenfunctions is constant,
\[
\langle\phi,\psi\rangle = 
\sum^{p}_{j=1}\left. \left[ \bar{\phi}_{n+j}
\psi^{\prime}_{n+j} -\bar{\phi}^{\prime}_{n+j} \psi_{n+j} \right]
\rule{0mm}{4.5mm}\right|^{a_j}_{0} = 0 .
\]
Here the two-form is the hermitian symplectic form in the space of
{\em boundary conditions} \cite{Kost:Sch}. The dimension of $N$ is
$2p$---there are two independent solutions for each edge. On the other
hand $N^{\perp}$ consists of the set of boundary conditions which match
on each of the interior edges but which may be arbitrarily prescribed on
the rays. \\  First let us consider $P\cap N^{\perp}$. These are boundary
conditions which `match' ($N^{\perp}$), as well as satisfy the
self-adjoint boundary conditions associated with $\cL$ ($P$).
Consequently, each element of 
$P\cap N^{\perp}$ can be identified with a generalised eigenfunction of
$\cL$ on the graph $\G$. However, these boundary conditions may also
describe solutions with support confined to the interior edges of the
graph, and we are only interested in solutions with support on the   
rays. \\
To pick only those solutions with suport on the rays we should consider
$N^{\perp}/N$. By lemma \ref{lfin} (in the appendix) this is a hermitian
symplectic space of dimension $2n$ and may be identified with the set of
boundary conditions with support on the rays. Consequently, if we
consider the projection of $P\cap N^{\perp}$ into $N^{\perp}/N$ we get
only those eigenfunctions with support on the rays and we know from
theorem \ref{th1} (in the appendix) that this space has dimension $n$.
\hspace*{\fill} $\Box$ \\

This, along with the fact that Lagrange planes are maximal 
isotropic subspaces, gives us the desired result.
\begin{cor}
Given the self adjoint extension $\cL$, the space of generalised 
eigenfunctions of $\cL$ at real eigenvalue
$\lambda$ and with support on the rays of the graph forms a
Lagrange plane in $H_{2n}(\lambda)$.
\end{cor}
Following Novikov we have an immediate application of these 
observations in the following proof of
the unitarity of the scattering matrix for $\lambda>0$ or real $k$. \\
Suppose that we have an $n$-dimensional basis for the space of generalised
eigenfunctions of the form\index{swg} 
\[
\swp{i} = \jsb{,i} + \sum_j S_{ij}\js{,j} .
\]
We call $S_{ij}$ the scattering matrix. Then since the generalised
eigenspace forms an isotropic subspace
\begin{eqnarray*}
0 & = & \langle \swp{i}, \swp{j} \rangle = \langle \jsb{,i} + \sum_l
S_{il}\js{,l},\; \jsb{,j} + \sum_m S_{jm}\js{,m} \rangle \\
& = & 2ik \left[
\sum_{l,m}\bar{S}_{il} S_{jm}\delta_{lm} - \delta_{ij} 
\right] ,
\end{eqnarray*}
where we have used equation (\ref{Jst}) for real $k$. Hence,
the scattering matrix is unitary for $\lambda>0$ or real 
$k$. \\ 
The original idea for this proof is to be found in corollary 2 of
\cite{Nov3} where the author uses it to prove
the {\em symmetry} of the scattering matrix (this is due
to the fact that Novikov uses {\em symplectic} geometry). \\
Similarly we can find a condition for the
symmetry of the scattering matrix. In the paper of Kostrykin and Schrader
\cite{Kost:Sch} the authors show that if the boundary conditions of an
operator can be expressed using real matrices then the scattering matrix is 
symmetric. In \cite{Har4} we show that all self-adjoint boundary
conditions can be parameterised by a unitary matrix $U$, the
condition of Kostrykin and Schrader's is equivalent to the symmetry
$U=U^T$ of $U$, which may also be written as the condition, $\phi\in
D(\cL) \Leftrightarrow \bar{\phi}\in D(\cL)$. Consequently the form
$\langle \swpb{i}, \swp{j}
\rangle$ is also zero
\begin{eqnarray*}
0 & = & \langle \swpb{i}, \swp{j} \rangle = \langle \js{,i} + \sum_l
\bar{S}_{il}\jsb{,l}, \jsb{,j} + \sum_m S_{jm}\js{,m} \rangle \\
& = & 2ik \left[ \sum_{m} S_{jm}\delta_{im} -  \sum_{l} S_{il} \delta_{lj} \right] 
\end{eqnarray*}
showing that the scattering matrix is symmetric. This is analogous to
Novikov's proof of the unitarity of the scattering matrix. \\

In the following sections we develop some new ideas based on Novikov's
construction. In particular, we show a link between the scattering matrix
and the Lagrange planes, and an application to the problem of
the factorisation of the scattering matrix. 

\section{The scattering matrix as parameter of the Lagrange planes}

We emphasise that for the remainder of this paper we will assume that
$\lambda>0$ or $k\in\R_0\equiv\R / \{ 0\}$. \\
We have shown that the space of generalised eigenfunctions
corresponds to a Lagrange plane, and that the Lagrange planes are
parameterised by a unitary matrix \cite{Har4}. It is not difficult to see 
that in the case of the asymptotic hermitian symplectic space this unitary
matrix is in fact the scattering matrix---for $\lambda>0$. First we
need some appropriate notation; we define a new hermitian symplectic form
simply by dividing the old form by $k$
\[
\langle\phi,\psi\rangle \equiv \frac{1}{k} \sum^{n}_{i=1}
\left[ \bar{\phi}_{i} \psi^{\prime}_{i}-\bar{\phi}^{\prime}_{i}
\psi_{i} \right] (x_i), \; x_i\in [0,\infty) .
\]
This is a hermitian symplectic form as long as $k$ is real and 
non-zero. In terms of this new form the Jost solutions
satisfy
\begin{equation}\label{Jsti}
\langle \js{,i},\js{,j}\rangle= 2i\delta_{ij} 
=-\langle \jsb{,i},\jsb{,j}\rangle \\
\langle \js{,i},\jsb{,j}\rangle = 0 
=\langle \jsb{,i},\js{,j}\rangle . 
\end{equation}
However, the canonical basis $\swp{0,i}$, $\swq{0,i}$ defined
above is not canonical anymore. Instead we define the new
canonical basis 
\begin{equation}\label{Can3}
\swp{0,j} = \frac{\js{,j}+\jsb{,j}}{2}, \qquad
\swq{0,j} = \frac{\js{,j}-\jsb{,j}}{2i}  
\end{equation}
where $j=1,\ldots ,n$. We also
use the notation
\[
\swz{0,j}=\swp{0,j}, \qquad \swz{0,j+n}=\swq{0,j} 
\]
where $j=1,\ldots ,n$ to denote these basis vectors. Let us denote by 
$\Lp{0,n}$ the Lagrange plane spanned by the first $n$
vectors of this basis. The precise relationship between the unitary
matrices and the Lagrange planes is given in corollary 2 of \cite{Har4}.
Here this result becomes:
\begin{thm}\label{Can5}
The Lagrange plane $\Lp{0,n}$ can be made to coincide with
$\Lp{n}$, the Lagrange plane corresponding to the generalised
eigenspace of a self-adjoint $\cL$, by means of the hermitian symplectic
transformation of the form
\begin{equation}
g=W^{\star}\hat{g}W=W^{\star}\left( \begin{array}{cc}
 S & 0 \\
 0 & \1
\end{array}
\right) W=\frac{1}{2}\left( \begin{array}{cc}
 S+\1 &  i(S-\1) \\
 -i(S-\1) &  S+\1
\end{array}
\right)
\end{equation}
where $S$ is the scattering\index{sm} matrix. \\
In particular, the canonical\index{cb} basis $\{\swz{0,i}\}^{2n}_{i=1}$ of
equation  (\ref{Can3}) is taken into a canonical basis
$\{\swz{i}\}^{2n}_{i=1}$
\[
\swz{i}=\sum^{2n}_{j}\,g_{ij}\swz{0,j} 
\]
where the first $n$ basis elements are the
scattering wave\index{swg} solutions of $\cL$ and so form a basis for
$\Lp{n}$.
\end{thm}
{\it Proof:} 
We substitute for $g$ and 
$\swz{0,i}$ to get for $i=1,\ldots ,n$
\begin{eqnarray*}
\swz{i} & = \frac{1}{2}\left[\sum^{n}_{j}(S+\1)_{ij} 
\left( \frac{\js{,j}+\jsb{,j}}{2} \right) + \sum^{n}_{j}i(S-\1)_{ij} 
\left( \frac{\js{,j}-\jsb{,j}}{2i} \right) \right] \\
& = \frac{1}{2}\left[\sum^{n}_{j}S_{ij}\js{,j}+\jsb{,i}\right] 
\equiv \swp{i} , 
\end{eqnarray*}
the scattering wave solution. \hspace*{\fill} $\Box$ \\

The remaining $n$ terms of
the new canonical basis, $\{\swz{i}\}^{2n}_{i=1}$, are denoted 
\[
\swq{i} = \swz{i+n} =
\frac{1}{2i}\left[\sum^{n}_{j}S_{ij}\js{,j}-\jsb{,i}\right] .
\]
Clearly this construction only works for $k\in\R_0$ when the
scattering matrix is unitary. In \cite{Har4} the matrix $U$ plays the
r\^{o}le of a unitary `parameter' which we were free to choose in order
to select self-adjoint boundary conditions and hence a Lagrange plane.
Here the unitary matrix valued function $S(k)$ of course depends in some
complicated way on the potentials on the edges and the boundary
conditions at the vertices. 

\section{The factorisation problem for the graph}

Suppose that we are given two non-compact graphs
$\G^{\prime}$ and $\G^{\prime\prime}$ with self-adjoint operators
$\cL^{\prime}$ and $\cL^{\prime\prime}$ defined on them and associated
scattering matrices $S^{\prime}$ and $S^{\prime\prime}$. Consider the
procedure of linking these graphs along $p$ of their (truncated) rays to
form a new graph $\G$. We can obviously define a self-adjoint operator on
$\G$ by using the boundary conditions and potentials of $\cL^{\prime}$ and
$\cL^{\prime\prime}$, we denote this by $\cL$. \\ 
Given $S^{\prime}$ and $S^{\prime\prime}$ and the details of the linking
can we find the scattering matrix $S$ of $\cL$? We will show that it is
possible to do so as long as the points at which rays are truncated in
order to form a linking edge are outside of the support of the
potential. \\

\subsection{Matching of asymptotic solutions on linking
edges}\label{match}

Consider a ray $r^{\prime}$ attached to $\G^{\prime}$ and
a ray $r^{\prime\prime}$ attached to $\G^{\prime\prime}$. We want to
connect these two rays together to form an edge of finite length in a new
graph $\G$. \\
We assume that the potentials on $r^{\prime}$ and
$r^{\prime\prime}$ have finite support; $\supp (q_{r^{\prime}})\subset
[0,a^{\prime}]$ and $\supp (q_{r^{\prime\prime}})\subset
[0,a^{\prime\prime}]$, respectively. We form the edges
$e^{\prime}=[0,a^{\prime}]$, $e^{\prime\prime}=[0,a^{\prime\prime}]$
by truncating the rays $r^{\prime}$, $r^{\prime\prime}$ at
$a^{\prime}$, $a^{\prime\prime}$, respectively and the two graphs
are linked simply by joining these edges end to end forming a new edge in
the interior of $\G$ of length $a^{\prime}+a^{\prime\prime}$. \\ 
\begin{defn}
Given $\psi_{\G^{\prime}}\in H_{2m^{\prime}}(\G^{\prime},\lambda)$ and 
$\psi_{\G^{\prime\prime}}\in 
H_{2m^{\prime\prime}}(\G^{\prime\prime},\lambda)$ we say that these
generalised eigenfunctions match on the edge
formed by joining $e^{\prime}$ and $e^{\prime\prime}$ end to end if
\begin{equation}\label{iso0}
\left.\psi_{\G^{\prime}}\rule{0mm}{4.5mm}\right|_{a^{\prime}} 
= \left.\psi_{\G^{\prime\prime}}\rule{0mm}{4.5mm}
\right|_{a^{\prime\prime}} \\
\left.\frac{d\psi_{\G^{\prime}}}{dx}\right|_{a^{\prime}} =
-\left.\frac{d\psi_{\G^{\prime\prime}}}{dx}\right|_{a^{\prime\prime}} 
\end{equation}
\end{defn}
That is the eigenfunctions $\psi_{\G^{\prime}}$,
$\psi_{\G^{\prime\prime}}$ match if together they represent a solution on
the augmented edge formed by joining
$e^{\prime}$ and $e^{\prime\prime}$ end to end---this is different from
the usage of the term `match' in Lemma 2 where instead of the asymptotic
hermitian symplectic space we were concerned with the space of boundary
values. Nevertheless, there are formal simularities between elements of
the the asymptotic hermitian symplectic space which match and elements of
the hermitian symplectic space of boundary values which match (although
there is no possibility of confusion as they are different spaces) which
is why we use the same term. \\

When considering linking edges it is natural to consider the 
sum
\[
H_{2m}(\lambda) = H_{2m^{\prime}}(\G^{\prime},\lambda) \oplus 
H_{2m^{\prime\prime}}(\G^{\prime\prime},\lambda)
\]
here $m=m^{\prime}+m^{\prime\prime}$. This is obviously also a hermitian
symplectic space with form
\[
\langle \phi_{\G^{\prime}}\oplus\phi_{\G^{\prime\prime}},\,
\psi_{\G^{\prime}}\oplus\psi_{\G^{\prime\prime}} \rangle \equiv
\langle \phi_{\G^{\prime}},\, \psi_{\G^{\prime}} \rangle_{\G^{\prime}} +
\langle \phi_{\G^{\prime\prime}},\, \psi_{\G^{\prime\prime}} 
\rangle_{\G^{\prime\prime}} 
\]
where $\langle\cdot,\cdot\rangle_{\G^{\prime}}$ and
$\langle\cdot,\cdot\rangle_{\G^{\prime\prime}}$ are the hermitian 
symplectic forms on $\G^{\prime}$ and $\G^{\prime\prime}$, respectively.
Using this notation 
the condition for matching is expressed in the
following lemma.
\begin{lem}\label{liso1}
The element 
\[
\psi = \psi_{\G^{\prime}}\oplus \psi_{\G^{\prime\prime}}\in H_{2m}
\] 
matches on the edge formed by joining $e^{\prime}$ and $e^{\prime\prime}$
iff
\begin{equation}\label{iso1}
\langle \psi_{\G^{\prime}}\oplus\psi_{\G^{\prime\prime}},\,
\zeta \js{,r^{\prime}}\oplus\jsb{,r^{\prime\prime}} \rangle = 0 \\
\langle \psi_{\G^{\prime}}\oplus\psi_{\G^{\prime\prime}},\,
\jsb{,r^{\prime}}\oplus\zeta \js{,r^{\prime\prime}} \rangle = 0 
\end{equation}
where $\zeta =e^{-ik(a^{\prime}+a^{\prime\prime})}$ and
$\jsd{,r^{\prime}}$ and $\jsd{,r^{\prime\prime}}$ are the Jost\index{jsg} 
solutions on the rays $r^{\prime}\in\G^{\prime}$ and
$r^{\prime\prime}\in \G^{\prime\prime}$, respectively.
\end{lem}
{\it Proof:} The Jost solutions $\jsd{,r^{\prime}}$ and 
$\jsd{,r^{\prime\prime}}$ form a basis on the rays $r^{\prime}$ and 
$r^{\prime\prime}$ so we can write
\[
\left.\psi_{\G^{\prime}}\rule{0mm}{4.5mm}\right|_{r^{\prime}} = 
\alpha^{\prime} \js{,r^{\prime}}+\beta^{\prime}\jsb{,r^{\prime}}, \qquad
\left.\psi_{\G^{\prime\prime}}\rule{0mm}{4.5mm}\right|_{r^{\prime\prime}}
= \alpha^{\prime\prime}\js{,r^{\prime\prime}}+
\beta^{\prime\prime}\jsb{,r^{\prime\prime}}
\]
Since the support of the potentials on the rays $r^{\prime}$ and 
$r^{\prime\prime}$ is within the intervals $[0,a^{\prime}]$ and
$[0,a^{\prime\prime}]$, and remembering that the Jost solutions are
continuous with continuous first derivatives, we see that 
\begin{eqnarray*}
\left.\jsd{,r^{\prime}}\rule{0mm}{4.5mm}\right|_{a^{\prime}} = 
e^{\pm ika^{\prime}}, & \qquad
\left.\frac{d\jsd{,r^{\prime}}}{dx}\right|_{a^{\prime}} =  
\pm ik e^{\pm ika^{\prime}} \\
\left.\jsd{,r^{\prime\prime}}\rule{0mm}{4.5mm}\right|_{a^{\prime\prime}}
= e^{\pm ika^{\prime\prime}}, & \qquad
\left.\frac{d\jsd{,r^{\prime\prime}}}{dx}\right|_{a^{\prime\prime}} =  
\pm ik e^{\pm ika^{\prime\prime}} .
\end{eqnarray*}
In order for equation (\ref{iso0}) to be satisfied we need the
following conditions
\begin{eqnarray*}
\alpha^{\prime}e^{ik a^{\prime}}+\beta^{\prime}e^{-ik a^{\prime}} & =
\alpha^{\prime\prime}e^{ik a^{\prime\prime}}+
\beta^{\prime\prime}e^{-ik a^{\prime\prime}} \\
\alpha^{\prime}e^{ik a^{\prime}}-\beta^{\prime}e^{-ik a^{\prime}} & =
- \left[ \alpha^{\prime\prime}e^{ik a^{\prime\prime}}-
\beta^{\prime\prime}e^{-ik a^{\prime\prime}} \right] , 
\end{eqnarray*}
or, solving for $\alpha^{\prime}$ and $\beta^{\prime}$,
\[
\bar{\zeta} \alpha^{\prime}=\beta^{\prime\prime}, \qquad
\beta^{\prime}=\bar{\zeta} \alpha^{\prime\prime} .
\]
On the other hand, using equation (\ref{Jsti}), we have
\begin{eqnarray*}
2i \bar{\alpha}^{\prime}=\langle\psi_{\G^{\prime}}, \js{,r^{\prime}} 
\rangle_{\G^{\prime}}, & \qquad
- 2i \bar{\beta}^{\prime}= \langle\psi_{\G^{\prime}}, \jsb{,r^{\prime}}
\rangle_{\G^{\prime}} \\
2i \bar{\alpha}^{\prime\prime}=\langle\psi_{\G^{\prime\prime}},
\js{,r^{\prime\prime}}\rangle_{\G^{\prime\prime}},
& \qquad
- 2i \bar{\beta}^{\prime\prime}= \langle\psi_{\G^{\prime\prime}},
\jsb{,r^{\prime\prime}}\rangle_{\G^{\prime\prime}} 
\end{eqnarray*}
so equation (\ref{iso0}) becomes
\begin{eqnarray*}
\zeta \langle\psi_{\G^{\prime}}, \js{,r^{\prime}} \rangle_{\G^{\prime}} 
& = - \langle\psi_{\G^{\prime\prime}}, \jsb{,r^{\prime\prime}}
\rangle_{\G^{\prime\prime}} \\
- \langle\psi_{\G^{\prime}}, \jsb{,r^{\prime}} \rangle_{\G^{\prime}} & = 
\zeta \langle\psi_{\G^{\prime\prime}},
\js{,r^{\prime\prime}}\rangle_{\G^{\prime\prime}}
\end{eqnarray*}
which, together with the fact that the hermitian symplectic form is 
linear in its second argument, gives the desired result. \hspace*{\fill}
$\Box$ \\
\begin{cor}\label{liso2}
The subspace of $H_{2m}$ of elements with support confined to the rays
$r^{\prime}$ and $r^{\prime\prime}$ and which match is an
isotropic\index{is} subspace with basis
\[
\left\{ \zeta \js{,r^{\prime}}\oplus\jsb{,r^{\prime\prime}},\, 
\jsb{,r^{\prime}}\oplus\zeta \js{,r^{\prime\prime}} \rule{0mm}{4.5mm}\right\}
\]
\end{cor}
{\it Proof:} The space of matching solutions is two-dimensional 
since this is simply the space of solutions on the augmented edge of
length $a^{\prime}+a^{\prime\prime}$. The
vectors we have given are independent---the Jost solutions $\js{}$ and
$\jsb{}$ are independent---all that remains is to show that they match,
which is easily seen to be true if they are put into equation
(\ref{iso1}). Furthermore, the fact that these basis vectors satisfy equation
(\ref{iso1}) means that they are contained in their orthogonal
complement, ie. the subspace is isotropic. \hspace*{\fill} $\Box$ \\ 

We consider linking $p$ of the rays of 
$\G^{\prime}$ with $p$ of the rays of $\G^{\prime\prime}$. Let us 
suppose that $\G^{\prime}$ has $m^{\prime}=n^{\prime}+p$ rays while 
$\G^{\prime\prime}$ has $m^{\prime\prime}=n^{\prime\prime}+p$ rays. We also
denote $m=m^{\prime}+m^{\prime\prime}$, $n=n^{\prime}+n^{\prime\prime}$
so that $m=n+2p$. \\ 
We choose $p$ of the rays of $\G^{\prime}$ and
$p$ of the rays of $\G^{\prime\prime}$ and consider the procedure of linking 
each ray of $\G^{\prime}$ with a ray of $\G^{\prime\prime}$ to form a
new graph $\G$. Let us denote by $N\subset H_{2m}$ the subset of elements
with support  confined to the linking rays and which match on the linking
rays. Then, by a simple generalisation of lemma \ref{liso2}, this
subspace is isotropic with dimension $2p$ and we can write a basis for it
in terms of the Jost\index{jsg} solutions on the linking rays similar to
the basis given in the lemma. \\ 
On the other hand, by lemma \ref{liso1} the elements $\psi\in H_{2m}$
which match on each of the linking rays are just those elements $\psi\perp
N$, ie. the subspace $N^{\perp}$. In summary, suppose we
choose $p$ rays, $\{ r^{\prime}_{i}\}^{p}_{i=1}$, of $\G^{\prime}$ and
$p$ rays, $\{ r^{\prime\prime}_{i}\}^{p}_{i=1}$, of $\G^{\prime\prime}$,
and consider linking $r^{\prime}_{i}$ to $r^{\prime\prime}_{i}$ for each
$i$ to form the graph $\G$. Then we have:
\begin{cor}\label{liso3}
The subspace $N\subset H_{2m}$ of elements with support confined to the
linking rays and which match on the linking rays is a $2p$-dimensional
isotropic\index{is} subspace with basis
\[
\left\{ \zeta_{i} \js{,r^{\prime}_{i}}\oplus\jsb{,r^{\prime\prime}_{i}},\, 
\jsb{,r^{\prime}_{i}}\oplus\zeta_{i} \js{,r^{\prime\prime}_{i}} 
\rule{0mm}{4.5mm}\right\}^{p}_{i=1} 
\]
where $\zeta_{i}=e^{-ik a_i }$ and $a_i$ is the length of the edge formed
by joining $r^{\prime}_{i}$ and $r^{\prime\prime}_{i}$.
Furthermore, $N^{\perp}\supset N$ consists of all of the elements of
$H_{2m}$ which match on the linking rays.
\end{cor}

\subsection{Description of the Lagrange plane of generalised
eigenfunctions for the linked graph $\G$}

We suppose that on the graphs $\G^{\prime}$ and $\G^{\prime\prime}$ we
have defined self-adjoint Schr\"{o}\-ding\-er operators $\cL^{\prime}$ and
$\cL^{\prime\prime}$ respectively. In terms of these operators we can
define the self-adjoint $\cL$ on the graph $\G$ formed by linking
$\G^{\prime}$ and $\G^{\prime\prime}$ as described above. \\ 
We recall that any generalised eigenfunction of $\cL$ can
be written as a generalised eigenfunction of $\cL^{\prime}$ on
$\G^{\prime}$ plus a generalised eigenfunction of $\cL^{\prime\prime}$ on
$\G^{\prime\prime}$ such that these two functions match on all of the
linking rays. This can be stated in the terms of the asymptotic hermitian
symplectic space: associated with $\cL^{\prime}$ and
$\cL^{\prime\prime}$ are the Lagrange planes
$\Lp{m^{\prime}}\subset H_{2m^{\prime}}(\G^{\prime})$ and 
$\Lp{m^{\prime\prime}}\subset
H_{2m^{\prime\prime}}(\G^{\prime\prime})$ respectively. Furthermore,
$\Lp{m}=\Lp{m^{\prime}}\oplus \Lp{m^{\prime\prime}}\subset H_{2m}$ is
a Lagrange plane. Then the intersection of
$\Lp{m}$ (the generalised eigenfunctions on $\G^{\prime}$ and 
$\G^{\prime\prime}$) and $N^{\perp}$ (the solutions which match on the
linking rays) gives us the generalised eigenfunctions of $\cL$ on $\G$. \\
Really we get a little bit more: $\Lp{m}\cap N^{\perp}$
may also contain solutions which have support only on the linking edges,
which, as we are only interested in solutions with support on the
semi-infinite rays, need to be discarded. 
In fact, we should not look for a solution in
the space $H_{2m}$ as it is not a suitable asymptotic hermitian symplectic
space for the linked graph $\G$. In particular, $H_{2m}$ has too high a 
dimension; the linked graph $\G$ has $n=n^{\prime}+n^{\prime\prime}$ rays so
we should be working in an asymptotic hermitian symplectic space of 
dimension $2n$. Consider the space $N^{\perp}/N$. It has dimension $2n$
(by lemma \ref{lfin}), moreover,  it consists of solutions that match on
all the linking rays. For this reason we state that $N^{\perp}/N$
is the asymptotic hermitian symplectic space for the linked graph
$\G$. \\
We have established that $\Lp{m}\cap N^{\perp}$ contains all of the 
generalised eigenfunctions of
the operator $\cL$ on the linked graph $\G$ plus, possibly, some
solutions with support on just the linking rays. Projecting 
$\Lp{m}\cap N^{\perp}$ onto $N^{\perp}/N$ eliminates solutions with
support only on the linking rays and so will give us the generalised 
eigenspace of $\cL$ on $\G$ which, by theorem \ref{th1} is a Lagrange
plane. 
\begin{cor}\label{thinf}
The Lagrange\index{Lp} plane $(\Lp{m}\cap N^{\perp})/N$ 
in $N^{\perp}/N$ corresponds to the space of generalised eigenfunctions 
with support on the rays for the operator $\cL$ on the graph $\G$.
\end{cor}
It is easy to see that this description generalises to
the case where an arbitrary number of graphs are linked. In this case
$\Lp{m}$ is defined as the direct sum of the Lagrange planes associated
with each of these graphs and $N$ is again an isotropic subspace which
describes how the graphs are to be linked. \\

\subsection{Description of the scattering matrix for the linked graph
$\G$}

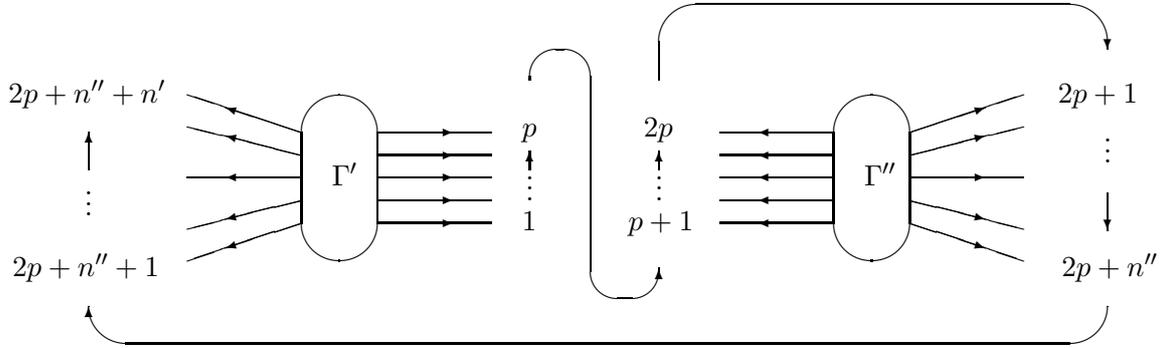
\begin{figure}[ht]{
\unitlength1mm\hspace*{-12mm}
\begin{picture}(140,60)
\put(45,30){\oval(10,22)}
\put(115,30){\oval(10,22)}
\put(44,29){$\G^\prime$}
\put(114,29){$\G^{\prime\prime}$}
\multiput(50,24)(0,3){5}{\line(1,0){15}}
\multiput(60,24)(0,3){5}{\vector(1,0){0}}
\multiput(110,24)(0,3){5}{\line(-1,0){15}}
\multiput(100,24)(0,3){5}{\vector(-1,0){0}}
\put(40,24){\line(-3,-1){15}}
\put(40,27){\line(-4,-1){15}} 
\put(40,30){\line(-1,0){15}}
\put(40,33){\line(-4,1){15}}
\put(40,36){\line(-3,1){15}}
\put(40,24){\vector(-3,-1){10}}
\put(40,27){\vector(-4,-1){10}} 
\put(40,30){\vector(-1,0){10}}
\put(40,33){\vector(-4,1){10}}
\put(40,36){\vector(-3,1){10}}
\put(120,24){\line(3,-1){15}}
\put(120,27){\line(4,-1){15}} 
\put(120,30){\line(1,0){15}}
\put(120,33){\line(4,1){15}}
\put(120,36){\line(3,1){15}}
\put(120,24){\vector(3,-1){10}}
\put(120,27){\vector(4,-1){10}} 
\put(120,30){\vector(1,0){10}}
\put(120,33){\vector(4,1){10}}
\put(120,36){\vector(3,1){10}}
\put(1.5,40){$2p+n^{\prime\prime}+n^{\prime}$}
\put(11.5,25){$\vdots$}
\put(12,31){\vector(0,1){5}}
\put(2,17){$2p+n^{\prime\prime}+1$}
\put(79,13){\oval(134,10)[b]}
\put(12,13){\vector(0,1){0}}
\put(69,35.5){$p$}
\put(69.5,27){$\vdots$}
\put(70,31){\vector(0,1){2.5}}
\put(69,23){$1$}
\put(140,17){$2p+n^{\prime\prime}$}
\put(145.5,32){$\vdots$}
\put(146,28){\vector(0,-1){5}}
\put(139.5,40){$2p+1$}
\put(146,48){\vector(0,-1){1}}
\put(116.5,48){\oval(59,10)[t]}
\put(87,43){\line(0,1){5}}
\put(85,35.5){$2p$}
\put(86.5,27){$\vdots$}
\put(87,31){\vector(0,1){2.5}}
\put(83,23){$p+1$}
\put(74,43){\oval(8,8)[t]}
\put(78,43){\line(0,-1){25}}
\put(82.5,18){\oval(9,8)[b]}
\put(87,18){\vector(0,1){0}}
\end{picture}}
\caption{\label{lnkg} Labeling of the rays of the graphs }
\end{figure}

For the sake of
convenience let us suppose that we are linking just two graphs $\G^{\prime}$
and $\G^{\prime\prime}$ (the case of an arbitrary number of graphs may
be reduced to this case). As
above, we assume that $\G^{\prime}$ has $m^{\prime}$ rays and
$\G^{\prime\prime}$ $m^{\prime\prime}$ rays and that we have selected $p$
rays of each graph to connect together. Consider the graph
$\G^{\prime}\oplus\G^{\prime\prime}$ and let us index the rays of
this graph according to the scheme
set out in figure \ref{lnkg}. \\

The first $p$ rays,
which are part of graph $\G^{\prime}$, are to be linked to the
next $p$ rays, which are part of graph $\G^{\prime\prime}$. The last $n$
rays in this scheme form the infinite rays of the linked graph $\G$, the
first $n^{\prime\prime}$ of these coming from $\G^{\prime\prime}$ and
the last $n^{\prime}$ coming from $\G^{\prime}$. \\
In order to make the calculations below clearer we introduce the
following index sets:
\begin{eqnarray*}
I & = & \{1,\ldots ,2m\} \\
I_N & = & \{1,\ldots ,2p\} \\
I_{N^{\perp}} & = & \{1,\ldots ,m,
2p+m+1, \ldots ,2m\} .
\end{eqnarray*}
We denote matrices in $\C^{m\times n}$ by $A_{(m,n)}$
and matrices in $\C^{n\times n}$ by $A_{(n)}$ where $\1_{(n)}$ is the
unit matrix in $\C^{n\times n}$. \\
The Jost solutions $\{\jsd{,i}\}^{m}_{i=1}$ and, as defined in equation
(\ref{Can3}), the canonical\index{cb} basis $\{\swz{0,j}\}^{2m}_{i=1}$ are
labelled in the obvious way according to the scheme of the figure. \\
We have self-adjoint $\cL^{\prime}$ and $\cL^{\prime\prime}$ defined
on $\G^{\prime}$ and $\G^{\prime\prime}$ respectively. Associated with
these operators we have the Lagrange planes $\Lp{m^{\prime}}$,
$\Lp{m^{\prime\prime}}$ and canonical bases as described in theorem
\ref{Can5}. Then
$\Lp{m^{\prime}}\oplus \Lp{m^{\prime\prime}}$ forms a Lagrange plane in
$H_{2m}$ with canonical basis $\{\swz{j}\}^{2m}_{j=1}$ inherited from the
canonical bases associated with $\Lp{m^{\prime}}$ and
$\Lp{m^{\prime\prime}}$. The indexing of the
basis elements $\{\swz{j}\}^{2m}_{j=1}$ 
follows the indexing given in figure \ref{lnkg}. Specifically, suppose
\begin{displaymath}
S^{\prime}_{(m^{\prime})}=\left( \begin{array}{cc}
 S^{\prime}_{(p)} & S^{\prime}_{(p,n^{\prime})} \\
 S^{\prime}_{(n^{\prime},p)} & S^{\prime}_{(n^{\prime})}
\end{array}
\right) 
\end{displaymath}
is the scattering matrix\index{sm} for $\cL^{\prime}$ and
\begin{displaymath}
S^{\prime\prime}_{(m^{\prime\prime})}=\left( \begin{array}{cc}
 S^{\prime\prime}_{(p)} & S^{\prime\prime}_{(p,n^{\prime\prime})} \\
 S^{\prime\prime}_{(n^{\prime\prime},p)} & 
S^{\prime\prime}_{(n^{\prime\prime})}
\end{array}
\right) 
\end{displaymath}
the scattering matrix\index{sm} for $\cL^{\prime\prime}$ where the {\em
ordering} of the entries follows the ordering described in
the figure---in particular the first
$p$ entries of each matrix correspond to the $p$ rays which are to be
linked. Then it is easy to see that the matrix $g$, which describes the
transformation from the basis $\{\swz{0,j}\}^{2m}_{i=1}$ to the basis
$\{\swz{j}\}^{2m}_{j=1}$ as in theorem \ref{Can5}, is of the form
\begin{displaymath}
g=W^{\star}\hat{g}W=W^{\star}\left( \begin{array}{cc}
 S_{(m)} & 0 \\
 0 & \1_{(m)}
\end{array}
\right) W
\end{displaymath}
where, following figure \ref{lnkg}
\begin{equation}\label{sm1}
S_{(m)}=\left( \begin{array}{cccc}
S^{\prime}_{(p)} & 0 & 0 & S^{\prime}_{(p,n^{\prime})} \\
0 & S^{\prime\prime}_{(p)} & S^{\prime\prime}_{(p,n^{\prime\prime})} & 0 \\
0 & S^{\prime\prime}_{(n^{\prime\prime},p)} & 
S^{\prime\prime}_{(n^{\prime\prime})} & 0 \\
S^{\prime}_{(n^{\prime},p)} & 0 & 0 & S^{\prime}_{(n^{\prime})}
\end{array}
\right) .
\end{equation}

We construct one more canonical\index{cb} basis,
$\{\swz{N,j}\}^{2m}_{j=1}$, which allows us to express the
isotropic\index{is} subspace $N$ in simple terms. Recalling corollary
\ref{liso3}, we see that the $2p$ elements defined by
\[
\swz{N,j} = \frac{\zeta_j \js{,j}+\jsb{,p+j}}{2}, \qquad
\swz{N,j+p} = \frac{\zeta_j \js{,p+j}+\jsb{,j}}{2}
\]
where $j=1,\ldots ,p$, $\zeta_j=e^{-ika_j}$ and $a_j$ is the length of the
$j$th linked edge, form a basis for $N$. Now we extend this to a
canonical basis by defining the following $n$
elements as identical to the elements of the canonical basis
$\{\swz{0,j}\}^{2m}_{j=1}$
\[
\swz{N,j}=\swz{0,j}=\frac{\js{,j}+\jsb{,j}}{2}
\]
where $j=2p+1,\ldots ,m$. Then these elements span a Lagrange
plane which, after theorem \ref{Can5}, has associated with it the
`scattering\index{sm} matrix'
\begin{equation}\label{sm2}
T_{(m)}=\left( \begin{array}{cc}
\begin{array}{cc}
0 & \zeta_{(p)} \\
\zeta_{(p)} & 0
\end{array}
& \mbox{\large{0}} \\
\mbox{\large{0}} & \1_{(n)}
\end{array}
\right)
\end{equation}
where $\zeta_{(p)}$ is a diagonal matrix with the entries on
the diagonal being the $\zeta_i$. It is then a simple matter to see that
the matrix
\[
g_N =W^{\star}\hat{g}_N W=W^{\star}\left( \begin{array}{cc}
 T_{(m)} & 0 \\
 0 & \1_{(m)}
\end{array}
\right) W
\]
takes the canonical basis $\{\swz{0,j}\}^{2m}_{j=1}$
into a new canonical basis $\{\swz{N,j}\}^{2m}_{j=1}$. We have already
shown that the linear span
\[
N=\bigvee_{j\in I_N} \{\swz{N,j}\}
\]
and it is not difficult to see, using the fact that this is a canonical
basis, that 
\[
N^{\perp}=\bigvee_{j\in I_{N^{\perp}} } \{ \swz{N,j}\} . 
\]

In order to get the scattering matrix\index{sm} for the linked
graph we first express the $\swz{j}$ in terms of the
$\swz{N,j}$. Since
\[
\swz{i}=\sum^{2m}_{j=1} g_{ij}\swz{0,j}, \qquad
\swz{N,i}=\sum^{2m}_{j=1} g_{N,ij}\swz{0,j}
\]
we can write
\begin{equation}\label{smh}
\swz{i}=\sum^{2m}_{j,k=1} g_{ij}g^{\star}_{N,jk}\swz{N,k}=
\sum^{2m}_{j=1} h_{ij}\swz{N,j}
\end{equation}
where
\begin{displaymath}
h =W^{\star}\hat{h}W=W^{\star}\left( \begin{array}{cc}
 S_{(m)} T^{\star}_{(m)} & 0 \\
 0 & \1_{(m)}
\end{array}
\right) W .
\end{displaymath}
We use this equation to find an $n$-dimensional canonical\index{cb}
basis for
$(\Lp{m}\cap N^{\perp})/N$. Clearly any such basis can be written, modulo
elements of $N$, as a linear combination of the
$\{\swz{j}\}^{m}_{j=1}$---since this set spans $\Lp{m}\supset (\Lp{m}\cap
N^{\perp})$. So there is a matrix
$R$ in $\C^{n\times m}$ so that the (representative in $\Lp{m}\cap
N^{\perp}$ of the) $i$-th basis element of
$(\Lp{m}\cap N^{\perp})/N$ is
\begin{equation}\label{sm3}
\sum^{m}_{j=1} R_{ij}\swz{j}=\sum^{m}_{j=1} \sum^{2m}_{l=1}
 R_{ij} h_{jl} \swz{N,l}
\end{equation}
Here $i=1,\ldots ,n$. What are the properties of the matrix $R$?
\begin{enumerate}
\item We can write the matrix $R$ in the form
\begin{equation}\label{sm5}
R=\left( \begin{array}{cc}
\rho_{(n,2p)} & \1_{(n)} \end{array} \right) .
\end{equation}
First we know that the subspace $N^{\perp}/N$ can be represented by
the space
\[
\bigvee_{j\in I_{N^{\perp}}\setminus I_N } \{ \swz{N,j}\} =
\bigvee_{j\in I_{N^{\perp}}\setminus I_N } \{ \swz{0,j}\} . 
\]
Furthermore, $(\Lp{m}\cap N^{\perp})/N$ is a Lagrange plane in this
space. Therefore, by theorem \ref{Can5}, there is a unitary matrix
$S_{(n)}$ such that the $i$-th basis element of the Lagrange plane has
the form 
\[
\frac{1}{2}\left[ \sum_j S_{(n),ij} \js{,j} + \jsb{,i} \right] .
\]
Here $i$ and $j$ take
values in the range $\{ 2p+1,\ldots ,m \}$. But this is equivalent to
equation (\ref{sm5}). Really, the only way to ensure that the $\jsb{,i}$
occur only `on the diagonal' is to have $\1_{(n)}$ in $R$, as shown.
\item In order for these basis elements to be in $\Lp{m}\cap N^{\perp}$
we need the coefficients of $\swz{N,l}$ for $l\in I\setminus
I_{N^{\perp}}$ in equation (\ref{sm3}) to be zero.
\end{enumerate}
Let us express the matrix $h$ in the following form:
\begin{equation}\label{sm4}
h = \left( \begin{array}{cc}
A_{(m)}  &  B_{(m)} \\
-B_{(m)} &  A_{(m)}
\end{array}
\right) = \left( \begin{array}{cccc}
A_{(2p)}    &  A_{(2p,n)} &   B_{(2p)} & B_{(2p,n)} \\
A_{(n,2p)}  &   A_{(n)}   & B_{(n,2p)} & B_{(n)} \\
-B_{(2p)}   & -B_{(2p,n)} &   A_{(2p)} & A_{(2p,n)} \\
-B_{(n,2p)} &   -B_{(n)}  & A_{(n,2p)} & A_{(n)}
\end{array}
\right)
\end{equation}
Using this representation, condition II can be expressed as
\[
\rho_{(n,2p)} B_{(2p)} + B_{(n,2p)}=0
\]
ie. 
\[ 
\rho_{(n,2p)}=-B_{(n,2p)}B^{-1}_{(2p)} .
\]
This gives us the matrix $R$ so we can write
\begin{eqnarray*}
R h  = \left( \begin{array}{ccc}
-B_{(n,2p)}B^{-1}_{(2p)}A_{(2p)}+A_{(n,2p)}, &
-B_{(n,2p)}B^{-1}_{(2p)}A_{(2p,n)}+A_{(n)},  &
0,    \end{array} \right. \\ 
 \qquad \left. \begin{array}{c} 
-B_{(n,2p)}B^{-1}_{(2p)}B_{(2p,n)}+B_{(n)} 
\end{array} \right) .
\end{eqnarray*}
We are not interested in the first $n\times 2p$ block of this
matrix as this represents the coefficients of the terms in $N$. Let us write
the second and fourth block as
$A$ and $B$, respectively. Then from condition I, along with theorem
\ref{Can5}, we see that 
\[
A = \frac{1}{2}(S_{(n)} + \1_{(n)}), \qquad
B = \frac{i}{2}(S_{(n)} - \1_{(n)})
\]
where, as above, $S_{(n)}$ is the desired scattering matrix of $\cL$ on
$\G$. In other words the scattering\index{fg}\index{sm} matrix is
\begin{eqnarray}\label{sm7}
S_{(n)} & = & A - i B \nonumber \\
& = & A_{(n)}-iB_{(n)}- B_{(n,2p)}B^{-1}_{(2p)}(A_{(2p,n)}-iB_{(2p,n)})
\nonumber \\ 
& = & \left( \begin{array}{cc}
S^{\prime\prime}_{(n^{\prime\prime})} & 0 \\ 0 &
S^{\prime}_{(n^{\prime})} \end{array} \right) +
\left( \begin{array}{cc} S^{\prime\prime}_{(n^{\prime\prime},p)}\bar{\zeta}_{(p)} & 0 \\
0 & S^{\prime}_{(n^{\prime},p)}\bar{\zeta}_{(p)} \end{array} \right)
\times \nonumber \\
& & \mbox{} \left(
\begin{array}{cc} \1{p} & -S^{\prime}_{(p)}\bar{\zeta}_{(p)} \\
-S^{\prime\prime}_{(p)}\bar{\zeta}_{(p)} & \1{p} \end{array} \right)^{-1}
\left( \begin{array}{cc} 0 & S^{\prime}_{(p,n^{\prime})} \\
S^{\prime\prime}_{(p,n^{\prime\prime})} & 0 \end{array} \right) \nonumber \\
& = & \left( \begin{array}{cc} S^{\prime\prime}_{(n^{\prime\prime})} & 0 \\
0 & S^{\prime}_{(n^{\prime})} \end{array} \right) +
\left( \begin{array}{cc} S^{\prime\prime}_{(n^{\prime\prime},p)} & 0 \\
0 & S^{\prime}_{(n^{\prime},p)} \end{array} \right)
\left( \begin{array}{cc} \zeta_{(p)} & -S^{\prime}_{(p)} \\
-S^{\prime\prime}_{(p)} & \zeta_{(p)} \end{array} \right)^{-1}
\nonumber \\  
& & \mbox{} \times \left( \begin{array}{cc} 0 &
S^{\prime}_{(p,n^{\prime})} \\ S^{\prime\prime}_{(p,n^{\prime\prime})} &
0 \end{array} \right) .
\end{eqnarray}

The inverse $B^{-1}_{(2p)}$ which appears above obviously may not always
exist. In fact $B_{(2p)}$ does not have an inverse
iff $\Lp{m}\cap N$ is non-empty---ie. iff we can find solutions with
no support on the external rays but support on the linking edges. \\
\begin{lem}
The matrix $B_{(2p)}$ does not have an inverse
iff $\Lp{m}\cap N$ is non-empty.
\end{lem}
{\it Proof:}
Let us suppose that $B_{(2p)}$ does not have an inverse. That is we can
find a non-zero vector $a$ such that
\[
a^{T} B_{(2p)} = 0 .
\]
Then, by equations (\ref{smh},\ref{sm4}), we get
\[
\psi = a^{T} \cdot \left( \begin{array}{c}
\swz{1} \\
\vdots \\
\swz{2p}
\end{array} \right) \in N^{\perp} .
\]
Now $\psi$ is clearly non-zero on the linking edges and has the form
$\alpha_i \js{,i}$ on the $n$ external rays---this statement follows
from the fact that the scattering waves $\swz{i}$ for $i=1,\ldots ,2p$
have this form on the $n$ external rays. Also $\psi$ is a generalised
eigenfunction for $\cL$, since it is in $\Lp{m}\cap N^{\perp}$. But then,
by theorem \ref{Can5}, all the $\alpha_i =0$. Another way to see this is
that, since $\psi$ is a generalised eigenfunction, it belongs to a Lagrange
plane and consequently
\[
\langle \psi , \psi \rangle = 0
\]
which is equivalent to all the $\alpha_i =0$. This gives us 
\[
\psi \in \Lp{m}\cap N \ne 0
\]
as required. \\
The converse statement follows simply from
\[
\swz{N,i} = \sum^{2m}_{j=1} h^{\star}_{ij}\swz{j} .
\]
This completes the proof. \hspace*{\fill} $\Box$ \\

This condition provides a means of identifying discrete\index{dee}
eigenvalues embedded in the continuous spectrum. 
\begin{cor}
Given a graph $\G$ with
$m$ vertices we split $\G$ into $m$ subgraphs
$\G_{d(1)},\ldots ,\G_{d(m)}$, each consisting of just one vertex with
$d(i)$ rays attach\-ed---here $d(i)$ is the degree of the $i$-th vertex of
$\G$. Then the
zeroes of the determinant of the matrix $B_{(2p)}$ for the set of
subgraphs
$\G_{d(1)},\ldots ,\G_{d(m)}$ give the discrete eigenvalues embedded in
the continuous spectrum.
\end{cor}
As we have mentioned above, equation (\ref{sm7}) for the scattering
matrix is the same, in essence, as the equation
given in the article by Kostrykin and Schrader \cite{Kost:Sch, Kost:Sch1}.
In this article the authors consider how the scattering matrix for the
Laplacean on a graph may be expressed in terms of the scattering matrices
of its subgraphs. Introducing a potential, as in the case of the
Schr\"{o}dinger operator, does not introduce anything essentially new (as
long as we assume, as we have done, that the potentials have compact
support and that we do not truncate rays inside the support).
Nevertheless our approach is sufficiently novel, we believe, to be of
independent interest. Kostrykin and Schrader also note the presence of an
inverse matrix in their formula (analogous to our matrix $B^{-1}_{(2p)}$)
and refer to the condition of this inverse not existing as {\it Condition
A}. \\

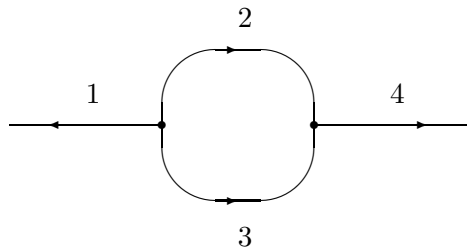
\begin{figure}[ht]
\centerline{
\unitlength1mm
\begin{picture}(60,60)
\put(20,30){\circle*{1}}
\put(40,30){\circle*{1}}
\put(0,30){\line(1,0){20}}
\put(40,30){\line(1,0){20}}
\put(30,30){\oval(20,20)}
\put(5,30){\vector(-1,0){0}}
\put(55,30){\vector(1,0){0}}
\put(10,33){1}
\put(50,33){4}
\put(30,40){\vector(1,0){0}}
\put(30,20){\vector(1,0){0}}
\put(30,14){3}
\put(30,43){2}
\end{picture}}
\caption{\label{figeg1} The graph from example  \protect\ref{eg1} }
\end{figure}

\begin{eg}\label{eg1}
Consider the graph in figure \ref{figeg1} with potential equal to zero on
all the edges, the internal edges of equal length $a$ and flux-conserved
boundary conditions, that is
\begin{eqnarray*}
\psi_1(0)=\psi_3(0)=\psi_4(0) \\
\psi_2(0)=\psi_3(a)=\psi_4(a) \\
\psi'_1(0)+\psi'_3(0)+\psi'_4(0)=0 \\
\psi'_2(0)-\psi'_3(a)-\psi'_4(a)=0 . 
\end{eqnarray*}
The arrows indicate the orientation of the edges.
\end{eg}

\begin{figure}[ht]
\centerline{
\unitlength1mm
\begin{picture}(60,60)
\put(30,30){\circle*{1}}
\put(10,30){\line(1,0){20}}
\put(30,30){\line(1,1){14}}
\put(30,30){\line(1,-1){14}}
\put(15,30){\vector(-1,0){0}}
\put(40,40){\vector(1,1){0}}
\put(40,20){\vector(1,-1){0}}
\end{picture}}
\caption{\label{Ygrph} The `Y' graph }
\end{figure}
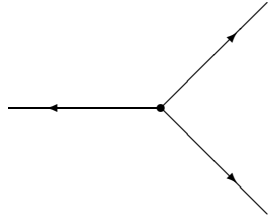

We can reconstruct the scattering matrix of the graph of figure \ref{figeg1}
by linking two `Y' graphs, depicted in figure \ref{Ygrph}, according to the
scheme of this section.

It is easy to see that the scattering matrices of such graphs have the form
\begin{displaymath}
S^{\prime}=S^{\prime\prime}=\left( \begin{array}{ccc}
-1/3 &  2/3 &  2/3 \\
 2/3 & -1/3 &  2/3 \\
 2/3 &  2/3 & -1/3
\end{array} \right) .
\end{displaymath}
So we get
\begin{eqnarray*}
& S^{\prime}_{(n^{\prime})} & =S^{\prime\prime}_{(n^{\prime\prime})}=
\left( \begin{array}{c} -1/3 \end{array} \right) \\
& S^{\prime}_{(n^{\prime},p)} & =S^{\prime\prime}_{(n^{\prime\prime},p)}=
\left( \begin{array}{cc} 2/3 & 2/3 \end{array} \right)=
S^{\prime T}_{(p,n^{\prime})}=S^{\prime\prime T}_{(p,n^{\prime\prime})} \\
& S^{\prime}_{(p)} & =S^{\prime\prime}_{(p)}=
\left( \begin{array}{cc} -1/3 & 2/3 \\
                          2/3 & -1/3 \end{array} \right)
\end{eqnarray*}
and, furthermore,
\begin{eqnarray*}
\left( \begin{array}{cc} S^{\prime\prime}_{(n^{\prime\prime})} & 0 \\
0 & S^{\prime}_{(n^{\prime})} \end{array} \right) & = &
\left( \begin{array}{cc} -1/3 & 0 \\
0 & -1/3 \end{array} \right) \\
\left( \begin{array}{cc} S^{\prime\prime}_{(n^{\prime\prime},p)} & 0 \\
0 & S^{\prime}_{(n^{\prime},p)} \end{array} \right) & = &
\left( \begin{array}{cccc}
2/3 & 2/3 & 0 & 0 \\
0 & 0 & 2/3 & 2/3 \end{array} \right) \\
\left( \begin{array}{cc} 0 & S^{\prime}_{(p,n^{\prime})} \\
S^{\prime\prime}_{(p,n^{\prime\prime})} & 0 \end{array} \right) & = &
\left( \begin{array}{cc}
0   & 2/3 \\
0   & 2/3 \\
2/3 &  0 \\
2/3 &  0
\end{array} \right) .
\end{eqnarray*}
Then from
\begin{displaymath}
\left( \begin{array}{cc} \zeta_{(p)} & -S^{\prime}_{(p)} \\
-S^{\prime\prime}_{(p)} & \zeta_{(p)} \end{array} \right) = 
\left( \begin{array}{cccc}
\zeta &   0   &  1/3  & -2/3 \\
   0  & \zeta & -2/3  & 1/3 \\
 1/3  & -2/3  & \zeta &  0 \\
-2/3  &  1/3  &   0   & \zeta
\end{array} \right) ,
\end{displaymath}
where of course $\zeta =e^{-ika}$, we can easily show that
\begin{eqnarray*}
\left( \begin{array}{cc} \zeta_{(p)} & -S^{\prime}_{(p)} \\
-S^{\prime\prime}_{(p)} & \zeta_{(p)} \end{array} \right)^{-1} =
\frac{\zeta}{(9\zeta^2 -1)(\zeta^2 -1)} \\
\mbox{} \times \left( \begin{array}{cccc}
9\zeta^2 -5 &   -4        & -(3\zeta+\bar{\zeta}) & 2(3\zeta-\bar{\zeta}) \\
    -4      & 9\zeta^2 -5 & 2(3\zeta-\bar{\zeta}) & -(3\zeta+\bar{\zeta}) \\
 -(3\zeta+\bar{\zeta}) & 2(3\zeta-\bar{\zeta}) & 9\zeta^2 -5 &   -4       \\
 2(3\zeta-\bar{\zeta}) & -(3\zeta+\bar{\zeta}) &     -4      & 9\zeta^2 -5
\end{array} \right) .
\end{eqnarray*}
Putting all of these into equation (\ref{sm7}) gives us the following
form for the scattering matrix
\[
S=\frac{1}{\gamma}\left( \begin{array}{cc}
3(\bar{\zeta}-\zeta) & 8 \\
8 & 3(\bar{\zeta}-\zeta)
\end{array} \right) 
\]
where we have used $\gamma =9\zeta-\bar{\zeta}$.

On the other hand, going back to the graph of figure \ref{figeg1}, it is easy
to see that this has a scattering wave solution of the form
\begin{eqnarray*}
\swp{1} & = e^{-ikx}+\frac{3(\bar{\zeta}-\zeta)}{\gamma}e^{ikx} \\
\swp{2/3} & = \frac{2\bar{\zeta}}{\gamma}e^{-ikx}+
\frac{6\zeta}{\gamma}e^{ikx} \\
\swp{4} & = \frac{8}{\gamma}e^{ikx} . 
\end{eqnarray*}
This confirms the form for the scattering matrix.

\section*{Acknowledgements}
The author would like to thank Prof B.S. Pavlov for his advice and
many useful conversations. 

\appendix
\section*{Appendix} 
\setcounter{section}{1}

Consider a canonical hermitian symplectic space $H_{2m}$ with Lagrange
plane $\Lp{m}$ and isotropic subspace $N$ of dimension $q$.
Then $N^{\perp}/N$ is a canonical hermitian symplectic space of dimension
$2n=2(m-q)$ and $(\Lp{m}\cap N^{\perp})$ projects to a Lagrange
plane in $N^{\perp}/N$.
\begin{lem}\label{lfin}
$N^{\perp}/N$ is a hermitian symplectic space of dimension 
$2n$
\end{lem}
{\it Proof:}  Since the form is nondegenerate
\[
\dim (N^{\perp}) = \dim (H_{2m})-\dim (N) = 2m - q = 2n + q
\]
Now since $N\subset N^{\perp}$; 
$\dim (N^{\perp}/N)=\dim (N^{\perp})-\dim (N)$ which gives us the 
result for the dimension. Clearly the form inherited from
$H_{2m}$ is uniquely defined since
\begin{displaymath}
\langle\phi + n_1 ,\psi + n_2\rangle=\langle\phi ,\psi \rangle
\qquad \phi ,\psi\in N^{\perp},\; n_1,n_2\in N
\end{displaymath}
To see nondegeneracy suppose there is some non-zero $\phi\in N^{\perp}$
which satisfies
\[
\langle\phi,\psi\rangle=0
\qquad \forall \psi\in N^{\perp} .
\]
But this simply means that $\phi\in N$, ie. $\phi$ is in the coset containing
zero. \hspace*{\fill} $\Box$ \\

Note that we have only shown that $N^{\perp}/N$ is a hermitian
symplectic space, not a canonical hermitian symplectic space. In order
to show that it is canonical we need only show (see \cite{Har4}) that it
contains a Lagrange plane:
\begin{thm}\label{th1}
The subspace $(\Lp{m}\cap N^{\perp})\subset N^{\perp}$ projects to a
Lagrange\index{Lp} plane in $N^{\perp}/N$.
\end{thm}
{\it Proof:} We denote, somewhat imprecisely, the projection of
$(\Lp{m}\cap N^{\perp})$ into $N^{\perp}/N$ by
$(\Lp{m}\cap N^{\perp})/N$. Clearly, $(\Lp{m}\cap N^{\perp})/N$ is
isotropic since $\Lp{m}$ is a Lagrange plane. We need only show that
it has maximal dimension. Firstly 
\begin{eqnarray*}
\dim ((\Lp{m}\cap N^{\perp})/N) & = \dim (\Lp{m}\cap N^{\perp}) -
\dim (\Lp{m}\cap N^{\perp}\cap N) \\
& = \dim (\Lp{m}\cap N^{\perp}) -
\dim (\Lp{m}\cap N) .
\end{eqnarray*}
Now, remembering that $\Lp{m}$ is a Lagrange
plane, it is easy to see that
\[
(\Lp{m}\cap N^{\perp})^{\perp} = \Lp{m} + N .
\]
So
\[
\dim (\Lp{m}\cap N^{\perp}) = \dim (H_{2m}) -
\dim (\Lp{m}+N) = 2m - \dim (\Lp{m}+N) .
\]
To proceed we use the basic vector space identity
$\dim (P)+\dim (N)=\dim (P+N)+\dim (P\cap N)$, which gives us
\[
\dim (\Lp{m}\cap N^{\perp}) =
2m-\left[ \dim (\Lp{m})+\dim (N)-
\dim (\Lp{m}\cap N) \right] .
\]
Putting this into our equation for
$\dim ((\Lp{m}\cap N^{\perp})/N)$ gives
\begin{eqnarray*}
\dim ((\Lp{m}\cap N^{\perp})/N) & =  
2m-\left[ \dim (\Lp{m}) + \dim (N) -
\dim (\Lp{m}\cap N) \right] - \mbox{} \\
 & \qquad \mbox{} - \dim (\Lp{m}\cap N) \\
 & = 2m - \dim (\Lp{m}) - \dim (N) \\
 & = m - q = n .
\end{eqnarray*}
This completes the proof. \hspace*{\fill} $\Box$ \\

Due to the nature of this proof it appears that this result should also
hold for symplectic geometry. We also note that these results are only
used above in the case $q=2p$ is even, although they clearly hold for any
integral $q$.


\begin{thebibliography}{1}

\bibitem{Har4}
M.~Harmer.
\newblock Hermitian symplectic geometry and extension theory.
\newblock {\em Journal of Physics A: Mathematical and General}, 33:9193--9203,
  2000.

\bibitem{Har}
M.~Harmer.
\newblock {\em The Matrix {Schr\"{o}dinger} Operator and {Schr\"{o}dinger}
  Operator on Graphs}.
\newblock PhD thesis, University of Auckland, 2000.

\bibitem{Kost:Sch}
V.~Kostrykin and R.~Schrader.
\newblock Kirchhoff's rule for quantum wires.
\newblock {\em J. Phys A: Math. Gen.}, 32:595--630, 1999.

\bibitem{Kost:Sch1}
V.~Kostrykin and R.~Schrader.
\newblock The generalized star product and the factorization of scattering
  matrices on graphs.
\newblock {\em J. Math. Phys}, 42:1563--1598, 2001.

\bibitem{Nov3}
S.~P. Novikov.
\newblock Schr{\"{o}}dinger operators on graphs and symplectic geometry.
\newblock In E.~Bierstone, B.~Khesin, A.~Khovanskii, and J.~Marsden, editors,
  {\em The Arnol'dfest (Toronto, ON, 1997)}, volume~24 of {\em Fields Institute
  Communications}, pages 397--413, 1999.

\bibitem{Pav}
B.~S. Pavlov.
\newblock The theory of extensions and explicitly solvable models.
\newblock {\em Uspekhi Math. Nauk-Russian. Math. Surveys}, 42(6):127--168,
  1987.

\end{thebibliography}
\end{document}